\documentclass[sigconf, review=false]{acmart}
\usepackage{array}
\usepackage{graphicx}
\usepackage{clrscode}
\usepackage{balance}
\usepackage{bbm}
\usepackage{subfigure}
\usepackage{multirow}
\usepackage{float}
\usepackage{color}
\usepackage{soul}
\usepackage{mdframed}
\usepackage{amsopn}
\usepackage{mathrsfs}
\usepackage{mathtools}
\usepackage{amsmath}
\usepackage{arydshln}
\usepackage{hyperref}
\usepackage{multicol}
\usepackage{blkarray}
\usepackage{enumerate}
\usepackage{enumitem}    
\usepackage{courier}
\usepackage{rotating}
\usepackage{booktabs}
\usepackage{diagbox}
\usepackage{fancybox}
\usepackage{minibox}
\usepackage{cases}
\usepackage{hhline}
\usepackage[lined,boxruled,commentsnumbered,linesnumbered]{algorithm2e}
\usepackage{adjustbox}
\usepackage{sistyle} 
\SIthousandsep{,}

\usepackage{xcolor}

\usepackage{changepage}

\copyrightyear{2020}
\acmYear{2020}
\setcopyright{acmlicensed}\acmConference[SIGIR '20]{Proceedings of the 43rd International ACM SIGIR Conference on Research and Development in Information Retrieval}{July 25--30, 2020}{Virtual Event, China}
\acmBooktitle{Proceedings of the 43rd International ACM SIGIR Conference on Research and Development in Information Retrieval (SIGIR '20), July 25--30, 2020, Virtual Event, China}
\acmPrice{15.00}
\acmDOI{10.1145/3397271.3401431}
\acmISBN{978-1-4503-8016-4/20/07}

\begin{document}
\fancyhead{}
\title{Understanding Echo Chambers in E-commerce\\ Recommender Systems}


\settopmatter{authorsperrow=4}

\author{Yingqiang Ge}
\affiliation{%
  \institution{Rutgers University}
}
\email{yingqiang.ge@rutgers.edu}
\authornote{Co-first authors with equal contributions.}
\authornote{This work was done when Yingqiang Ge worked as an intern in Alibaba.}

\author{Shuya Zhao}
\affiliation{%
  \institution{New York University}
}
\email{sz2257@nyu.edu}
\authornotemark[1]

\author{Honglu Zhou}
\affiliation{%
  \institution{Rutgers University}
}
\email{honglu.zhou@rutgers.edu}

\author{Changhua Pei}
\affiliation{%
  \institution{Alibaba Group}
}
\email{changhuapei@gmail.com}

\author{Fei Sun}
\affiliation{%
  \institution{Alibaba Group}
}
\email{ofey.sunfei@gmail.com}

\author{Wenwu Ou}
\affiliation{%
  \institution{Alibaba Group}
}
\email{santong.oww@taobao.com}

\author{Yongfeng Zhang}
\affiliation{%
  \institution{Rutgers University}
}
\email{yongfeng.zhang@rutgers.edu}








\begin{abstract}
Personalized recommendation benefits users in accessing contents of interests effectively. 
Current research on recommender systems mostly focuses on matching users with proper items based on user interests. 
However, significant efforts are missing to understand how the recommendations influence user preferences and behaviors, e.g., if and how recommendations result in \textit{echo chambers}. 
Extensive efforts have been made in examining the phenomenon in online media and social network systems.
Meanwhile, there are growing concerns that recommender systems might lead to the self-reinforcing of user's interests due to narrowed exposure of items, which may be the potential cause of echo chamber. 
In this paper, we aim to analyze the echo chamber phenomenon in Alibaba Taobao --- one of the largest e-commerce platforms in the world.

Echo chamber means the effect of user interests being reinforced through repeated exposure to similar contents. 
Based on the definition, we examine the presence of echo chamber in two steps. First, we explore whether user interests have been reinforced. 
Second, we check whether the reinforcement results from the exposure of similar contents. 
Our evaluations are enhanced with robust metrics, including cluster validity and statistical significance.
Experiments are performed on extensive collections of real-world data consisting of user clicks, purchases, and browse logs from Alibaba Taobao.
Evidence suggests the tendency of echo chamber in user click behaviors, while it is relatively mitigated in user purchase behaviors. 
Insights from the results guide the refinement of recommendation algorithms in real-world e-commerce systems.
\end{abstract}



\maketitle

\section{Introduction}
\label{sec:introduction}
Recommender systems (RS) comes into play with the rise of online platforms, e.g., social networking sites, online media, and e-commerce\cite{ge2019maximizing, ge2020learning, FuXianGaoZhaoHuangGeXuGengShahZhangDeMelo2020FairRecommendation}. 
Intelligent algorithms with the ability to offer personalized recommendations are increasingly used to help consumers seek contents that best match their needs and preferences in forms of products, news, services, and even friends\cite{zhang2013localized,allen2017effects,zheng2018drn}. 
Despite the significant convenience that RS has brought, the outcome of the personalized recommendations, especially how it reforms social mentality and public recognition --- which could potentially reconfigure the society, politics, labor, and ethics --- remains unclear. 
Extensive attention has been drawn at this front, thus arriving at the two coined terms, \textit{echo chamber} and \textit{filter bubble}. 
Both effects might occur after the use of personalized recommenders and entail far-reaching implications. 
Echo chamber describes the rising up of social communities who share similar opinions within the group~\cite{sunstein2009going}, while filter bubble~\cite{pariser2011filter}, as the phenomenon of an overly narrow set of recommenders, was blamed for isolating users in information echo chambers~\cite{allen2017effects}.

Owing to the irreversible and striking impact that the internet has brought on the mass communication, echo chamber and filter bubble are appearing in online media and social networking sites, such as MovieLens~\cite{nguyen2014exploring}, Pandora~\cite{allen2017effects}, YouTube~\cite{hilbert2018communicating}, Facebook~\cite{risius2019towards}, and Instagram~\cite{stoica2019hegemony}. 
Significant research efforts have been put forward in examining the two phenomena in online media and social networks~\cite{geschke2019triple,bountouridis2019siren,mohseni2018combating,burbach2019bubble,flaxman2016filter,bakshy2015exposure}. 
Recently, researchers have concluded that the decisions made by RS can influence user beliefs and preferences, which in turn affect the user feedback, e.g., the behavior of click and purchase received by the learning system, and this kind of user feedback loop might lead to echo chamber and filter bubbles~\cite{jiang2019degenerate}. 
On the other hand, the two concepts are not isolated, since filter bubble is a potential cause of echo chamber~\cite{allen2017effects,dash2019network}.

In this work, we are primarily concerned with the existence and the characteristics of echo chamber in real-world e-commerce systems. 
We define echo chamber as the effect of users' interests being reinforced due to repeated exposure to similar items or categories of items, thereby generalizing the definition in~\cite{jiang2019degenerate}. 
This is because users' consuming preferences are so versatile and diverse that cannot simply be classified into positive or negative directions as what it looks like in political opinions~\cite{10.2307/j.ctt7tbsw}.
Based on the above definition of echo chamber, we formulate the research in two steps by answering the following two related research questions:
\begin{itemize}
    \item RQ1: Does the recommender system, to some extent, reinforce user click/purchase interests?
    \item RQ2: If user interests are indeed strengthened, is it caused by RS narrowing down the scope of items exposed to users?
\end{itemize}

To measure the effect of recommender systems on users, we first follow the idea introduced in ~\cite{nguyen2014exploring} and separate all users into categories based on how often they actually ``take'' the recommended items. 
This separation helps us to compare recommendation followers against a controlled group, namely, the recommendation ignorers.
The remaining problem is how to measure the effect of echo chamber on each group.
Users in social network platforms have direct ways to interact with other users, potentially through actions of friending, following, commenting, etc~\cite{sasahara2019inevitability}. 
A similar analogy is that users in the recommender system could interact with other users \textit{indirectly} through the recommendations offered by the platform, since recommendation lists are usually generated as a result of considering the user's previous preferences and the preferences of similar users ($i.e.$, collaborative filtering). 
Due to the absence of an explicit network of user-user interaction, which is naturally and commonly provided in social networks, we decide to measure echo chamber in e-commerce at the population level.
This is because users who share similar interaction records ($e.g.,$ clicking the same products) will be closely located in a latent space, and the cluster of these users in that space, along with its temporal changes, could serve as signals to detect echo chamber. 
Finally, we measure the content diversity in recommendation lists for each group to see whether recommender system narrows down the scope of items exposed to users, so as to answer RQ2.

The key contributions of our paper can be summarized as follows:
\vspace{-15pt}
\begin{itemize}
\item We study echo chamber effect at a population level by implementing clustering on different user groups, and measure the shifts in user interests with cluster validity indexes.

\item We design a set of controlled trials between recommendation followers and ignorers, and employ a wide range of technical metrics to measure the echo chamber effect to provide a broader picture.

\item  We conduct our experiments based on real-world data from Alibaba Taobao --- one of the largest e-commerce platforms in the world.
Our analytical results, grounded with reliable validity metrics, suggest the tendency of echo chamber in terms of the user click behaviors, and relatively mitigated effect in user purchase behaviors.
\end{itemize}


\section{Related work} \label{sec:related}
Today's recommender systems are criticized for bringing dangerous byproducts of echo chamber and filter bubble. 
Sunstein argued that personalized recommenders would fragment users, making like-minded users aggregate \cite{sunstein2009going}. 
The existing views or interests of these users would be reinforced and amplified since ``group polarization often occurs because people are telling one another what they know''~\cite{sunstein2009going,sunstein2018republic}. Pariser later described filter bubble, as the effect of recommenders making users isolated from diverse content, and trapping them in an unchanging environment~\cite{pariser2011filter}. Though both are concerned with the malicious effect that recommenders would pose, echo chamber emphasizes the polarized environment, while filter bubble lays stress on the undiversified environment.

Researchers are expressing their concerns of the two effects, and attempting to formulate a richer understanding of the potential characteristics~\cite{stoica2019hegemony,cohen2018exploring,namjun2019effect,hilbert2018communicating,bakshy2015exposure}. Considering echo chamber as a significant threat to modern society as they might lead to polarization and radicalization~\cite{dandekar2013biased}, Risius et al. analyzed news ``likes'' on Facebook, and distinguished different types of echo chambers~\cite{risius2019towards}. 
Mohseni et al. reviewed news feed algorithms as well as methods for fake news detection and focused on the unwanted outcomes of echo chamber and filter bubble after using personalized content selection algorithms~\cite{mohseni2018combating}. They argued that personalized newsfeed might cause polarized social media and the spread of fake content. 

Another genre of research aims to clear up strategies to mitigate the potential issues of echo chamber and filter bubble, or design new recommenders to alleviate such effects~\cite{badami2018prcp,antikacioglu2017post,
helberger2018exposure,de2014emotions,pardos2019combating,gao2018burst}. 
Badami et al. proposed a new recommendation model for combating over-specialization in polarized environments after finding that matrix factorization models are easier to learn in polarized environments, and in turn, encourage filter bubbles that reinforce polarization. Tintarev el al. attempted to use visual explanations, i.e., chord diagrams and bar charts, to address the problems~\cite{tintarev2018knowing}. 

There is a certain amount of work focusing on the detection or measuring of echo chamber and filter bubble, questioning whether they do exist~\cite{allen2017effects,fleder2007recommender,bakshy2015exposure,nguyen2014exploring,jiang2018community,moller2018not}. For example, Hosanagar et al. used data from an online music service, trying to find out whether personalization is, in fact, fragmenting the population, and concluded that it does not~\cite{hosanagar2013will}. They claimed personalization is a tool that helps users widen their interests, which in turn creates commonality with others. 
Sasahara et al. suggested echo chambers are somewhat inevitable given the mechanisms at play in social media, specifically, the basic influence and unfriending~\cite{sasahara2019inevitability}. Their simulation dynamics showed that the social network rapidly devolves into segregated, homogeneous, and polarized communities, even with the minimal amount of influence and unfriending.  

Despite the reasonableness of prior works, severe limitations do exist, making the claims only plausible. One major aspect is that most of the existing works draw conclusions by means of simulation, or relying on some self-defined networks and measurements with simplified dynamics~\cite{geschke2019triple,bountouridis2019siren,chaney18algorithmic,jiang2019degenerate,fleder2007recommender,moller2018not}. Building upon the subjective assumptions, whether the modeling and analysis have the capability to reflect the truth seems to be dubious~\cite{risius2019towards}. On the other hand, many of the prior works confound the meaning of the two effects or solely examine one of them without considerations of the other~\cite{jiang2018community,gao2018burst,hilbert2018communicating,kim2017echo,risius2019towards}. Exceptions such as~\cite{jiang2019degenerate}, disentangles echo chamber from filter bubble, but suffers from the previous-mentioned deficiency, i.e., reliance on simulation and simplified artificial settings. With the desire to address the limitations, we aim to explore the existence of echo chamber in real-world e-commerce systems while investigating filter bubble via differentiating it as the potential cause of echo chamber. 
To the best of our knowledge, this is the first work that utilizes real-world data of recommendation and user-item interaction from a large-scale real-world e-commerce platform, with solid validity metrics, instead of from the artificial well-controlled experimental settings. We do not induce any prior assumptions that might be unreliable. We draw analysis from a latent space without relying on any explicit pre-built networks.





\vspace{-5pt}
\section{Data Collection and Analysis}
\label{sec:data_prep}
\subsection{Data Collection}
\label{sec:data_collection}
We first collected $86,192$ users' five months accessing logs spanning Jan. 1, 2019 to May 31, 2019, from Taobao.
There exist three types of accessing logs: browse log, click log, and purchase log. 

\textbf{Browse Log} records the recommendations that have been browsed for each user, including the timestamp, page view id, user id, recommended item id, item position in the recommendation list, and whether it was clicked or not.
The recommended items per page are known as a page view (PV) in e-commerce. 

\textbf{Click Log} is used to record each user's click behaviors in the whole electronic market, which means that user click behaviors outside the recommender list will also be recorded (i.e., the user may initiate a search and click an item).
It includes the timestamp, PV id, user id, user profile, clicked item id, and the item price.

\textbf{Purchase Log} 
is in the same format as the click log, except it is used to record users' purchase behaviors. Besides, since purchase is a more expensive action than click for users, the sparsity of the purchase log is much higher than that of the click log.

We extract dataset with the above components (browse log, click log, and purchase log) for the following three reasons:
\vspace{0pt}
\begin{itemize}
\item E-commerce is a complicated scenario, 
where users have multiple types of interactions with items, such as browsing, clicking, purchasing, and rating.
The effect of RS might have a different influence on them, because different interactions induce different costs for users, e.g., browsing and clicking merely cost time, while purchasing costs time and money.

\item Clicking and purchasing are used to represent consumer's implicit feedback in RS, which contains rich information about consumer preferences. 
However, browsing contains much more noise as most of the browsed items may be indifferent to a user, which means the user neither likes nor dislikes the item. 

\item In RS, ratings are user's explicit feedback to items.
However, since the platform automatically fulfills the rating score to 5 stars (the highest score) if a user purchased an item without leaving a rating, so the ratings may not reflect users' true preference, thus, we do not use ratings in this work.
\end{itemize}

We further remove each user's logs in the first two months to make sure that all users have enough time to get familiar with the e-commerce platform, and that the RS receives sufficient interactions from consumers to understand their preferences well.

\begin{figure}[]
\vspace{-15pt}
\centering
\mbox{
\hspace{-10pt}
\centering
    \label{fig:value1}
    \includegraphics[scale=0.45]{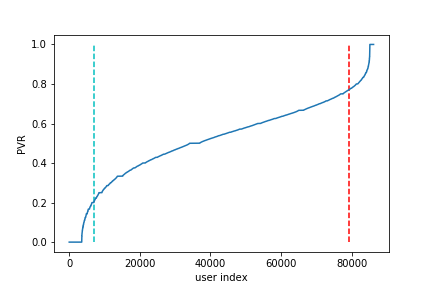}
}
\vspace{-10pt}
\caption{The users sorted by PVR from the lowest to the highest score. 
Each x-axis index refers to a unique user.
The two split points represent 20\% (blue) and 80\% (red) users.}
\label{fig:PVR}
\vspace{-18pt}
\end{figure}

\subsection{Identifying Recommendation Takers}
The purpose of our work is to explore the temporal effect of recommendation systems on users, especially to investigate the existence and the characteristics of the echo chamber effect.
In order to study the effect of ``taking'' recommendations, we need to classify the users in the dataset into users who ``follow'' recommendations and users who do not.
For consistency, in this paper, we call the ones who ``follow'' recommendations as the \textit{Following Group} and those who do not as the \textit{Ignoring Group}.
Inspired by the experiment setting of \cite{nguyen2014exploring}, we draw comparisons between two groups of consumers --- the \textit{Following Group} and the \textit{Ignoring Group}--- to explore the effect of recommendation systems on users.


The first step is to classify users into the two groups, and there are several different approaches to this classification. 
One straightforward approach is to use the ratio between the number of clicked items and the number of browsed items. We can calculate this ratio for each user $i$ based on his or her browsing history. 
However, one extreme case is that user $i$ only viewed each item once, and the user clicked most of them, but never 
came back to the platform to use the recommendation system again.
This will classify the user into the \textit{Following Group} because the ratio is close to one but he/she is misclassified because he/she actually abandoned the recommender system.
We can see that this approach cannot help to investigate the long-term influence of RS on consumers.

In order to fulfill the need for long-term observation, we adopt the ``block'' idea \cite{nguyen2014exploring} into our classification task.
We first identify clicked PV, which is a recommendation list on a single page 
where the consumer clicked at least one item displayed in it. 
Then, we compute the number of clicked PVs over the total number of all PVs for a given consumer, and we define this ratio as PVR (namely, page view ratio).
The intuition behind this design is that we believe the effect that RS imposes on consumers depends on the frequency that consumers are exposed to it, as well as consumers' responses to the recommendations ($e.g.$, clicks).
Once we have calculated the PVRs for all users, we sort the users from the lowest to the highest PVR scores, and the result is shown in Figure~\ref{fig:PVR}. 
Based on this figure, we define users who took recommendations in at most 20\% of their PVs as the \textit{Ignoring Group}, and users who took recommendations in at least 80\% of their PVs as the \textit{Following Group}, which gives us 6,183 followers and 6,979 ignorers in total.

\subsection{User Interaction Blocks}
In order to examine the temporal effect of recommender systems on users, we divide the user-item interaction history (could be $browse$, $click$, or $purchase$) into discrete intervals for each user.
We follow the similar ``blocks of interaction'' idea in \cite{nguyen2014exploring} to divide the interaction history of a user into intervals, which is used to make sure that all users have the same amount of interactions with the recommender system throughout an interaction block.

We define an interval as a block consisting of $n$ consecutive interactions, where $n$ is a constant decided by experimental studies.
Moreover, different interactions occur at different frequencies, for example, the number of browsed items is much higher than the number of clicked items, and the number of clicked items is again higher than the purchased items, indicating that the length of the block (i.e., $n$) may vary based on the corresponding interactions.
We primarily set the length of the interval as $n=200$ for browsed items (named as a browsing block), $n=100$ for clicked items (named as a clicking block), and $n=10$ for purchased items (named as a purchasing block).
If there are not enough interactions to constitute the last block, we will drop it to make sure that all blocks have the same number of interactions.
Meanwhile, to ensure the temporal effect of RS, we only keep those users who have at least three intervals in the three months, for each type of user-item interaction.
Finally, as shown in Table~\ref{table:users}, we have 7,477 users to examine the echo chamber effect on click behaviors, 3,557 users for purchase behaviors, and 7,417 users for browsing behaviors.

Considering the browse log is potentially noisy (i.e., indifferent to a user, see Section~\ref{sec:data_collection}), as well as the fact that clicking and purchasing are commonly used to represent users' implicit feedback to RS, in the following, we use click and purchase behaviors to represent users' preferences on the items (while we detect the echo chamber effect), and examine the temporal changes in content diversity of recommended items via browsing behaviors (which may be the potential cause of echo chamber).
\vspace{-5pt}

\subsection{User Embeddings}
\label{sec:embedding}
As the user interests are closely related to the user interactions with items, we argue that the items that the user clicked can reflect his/her click interests, and the items that the user purchased can represent his/her purchase interests.
However, only using discrete indexes to denote the interacted items is not sufficient to represent user interests since we cannot know the collaborative relations between different items only based on the indexes. 
Following the basic idea of collaborative filtering, we use the user-item interaction information to train an embedding model. 

Items are encoded into item embeddings based on one of the state-of-the-art models \cite{DBLP:journals/corr/abs-1803-02349}.
To cluster and compare the items for different users at different times, we need to guarantee that the item embeddings are stable across the period of time that is under investigation. For this purpose, the embeddings are trained on all of the collected data until May 31, 2019, which is the last day that our dataset contains.
This is for two reasons: (1) since the training data contains all of the user-item interactions, it helps to learn more accurate embeddings; and (2) since the training procedure includes all items under consideration, we can guarantee that all embeddings are learned in the same space.
After that, we use the average pooling on the item embeddings to compute the user embeddings. 
Specifically, we use the average of the item embeddings of items that the user clicked (or purchased) within a user-item interaction block 
to represent the user's click preferences (or purchase preferences) during a certain period of time.
In this way, user embeddings and item embeddings are in the same representation space.

\begin{table}[H]
\small
\centering
\renewcommand{\multirowsetup}{\centering}
\begin{tabular}{cccc}
\toprule
& Click & Purchase& Browse \\ \midrule
Following group  & $5,025$  & $2,099$ & $5,507$\\ \midrule
Ignoring group & $2,452$  & $1,458$  & $1,910$ \\ \midrule
All users  & $7,477$  &  $3,557$  & $7,417$ \\ \bottomrule 
\end{tabular} 
\caption{Statistics of each user group.}
\label{table:users}
\vspace{-20pt}
\end{table}

\section{Measures for Echo Chambers}
\label{sec:metric}

To answer RQ$1$, we propose to study the reinforcement in user interests at a population level. The pattern of reinforcement could happen in a simple scenario, where some members highly support one opinion, while others believe in a competing opinion. The phenomenon is reflected as the dense distribution on the two sides of the opinion axis. However, user's interest in e-commerce is much more complicated, such that it cannot be simply classified into positive and negative. What we observe is that users can congregate into multiple groups in terms of distinct preferences. As a result, we implement clustering on user embeddings and measure the change in user interests with cluster validity indexes
(more details can be found in Section~\ref{sec:user_rein}).
We measure the changes in terms of clustering on the embeddings at the beginning and at the end of the user interaction record, i.e., we compute the user embeddings respectively for the first and the last interaction block and measure the changes. To be clear, we refer to these two blocks as the ``first block'' and the ``last block''.

To answer RQ$2$, we propose to measure the content diversity in recommendation lists at the beginning and the end, and more details can be found in Section~\ref{sec:metric_diversity}. We examine whether there exists a trend that recommendation systems narrow down the contents provided to the users.
Before we cluster the \textit{Following Group} and the \textit{Ignoring Group}, respectively, we need to know whether the two groups are clusterable and what is the appropriate number of clusters for each of the group. Thus, we first examine the clustering tendency and select the proper clustering settings, which will be introduced in Section~\ref{sec:settings}.

\subsection{Measuring Clusters}
\label{sec:settings}


\subsubsection{Clustering Tendency}
\mbox{} \\
Assessing clustering tendency is employed to evaluate whether there exist meaningful clusters in the dataset before applying clustering methods. We use \textbf{Hopkins statistic} (H) \cite{lawson1990new, banerjee2004validating} to measure the tendency since it can examine the spatial randomness of the data by testing the given dataset with a uniformly random-distributed dataset. The value of H is from $0$ to $1$. A result close to $1$ indicates a highly clustered dataset, while a result around $0.5$ indicates that the data is random.

Let $X \in R^{D}$ be the given dataset of $N$ elements, and $Y \in R^{D}$ is the uniformly random dataset of $M$ ($M\ll N$) elements with the same variation as $X$. Then we get a random sample \{$x_{1}^{D},x_{2}^{D},\dots,x_{M}^{D}$\} from X. And $s_{i}^{D}$ and $t_{i}^{D}$ are the distances from $x_{i}^{D}$ and $y_{i}^{D}$ to their nearest neighbor in $X$, respectively.
Hopkins statistic is computed as the following:
\begin{equation}
\small
H={\sum\nolimits_{i=1}^{M}t_{i}^{D} \over \sum\nolimits_{i=1}^{M}s_{i}^{D}+\sum\nolimits_{i=1}^{M}t_{i}^{D}}
\end{equation} 

The results are shown in Table~\ref{table:hopkins}. 
Hopkins statistic 
examines the datasets before applying further measurement to them. 
The characteristics of a clusterable dataset ($H>0.5$) can be observed under its optimal setting in K-means clustering. Then we can select the proper number of clusters for each user group.

\subsubsection{Clustering Settings}
\mbox{} \\
We use the \textbf{Bayesian Information Criterion} (BIC) to determine the number of clusters for each group. 
Due to the high-dimensional characteristics of user embeddings, it is hard to choose the optimal $K$ (i.e., the number of clusters) via some common k-selection techniques, like elbow method or average silhouette method. 
To deal with it, we use model selection technique to compare the clustering results under different $K$s and we choose BIC, which aims to select the model with maximum likelihood. 
Its revised formula for partition-based clustering suits our tasks well and there is also a penalty term avoiding overfitting in the formula.

\begin{equation}
\Small{
\begin{aligned}
BIC= &\sum\limits_{i=1}^{K} n_{i}\Bigl( \log{n_{i} \over N}-\frac{n_{i}D\log{2\pi \Sigma}}{2} - \frac{D(n_{i}-1)}{2}\Bigr)  -\frac{K(D+1)\log{N}}{2} \\
\end{aligned}
}
\end{equation}
where the variance is defined as $\Sigma =\frac{1}{N-K} \sum\limits_{i=1}^{K}\sum\limits_{j=1}^{n_{i}}\bigl\Vert x_{j}-c_{i}\bigr\Vert_{2}$.

The $K$-class clustering has $N$ points $x_{j} \in X^{D}$, $c_{i}$ is the center of the $i$-th cluster with the size of $n_i$, $i=1,\dots,K$. BIC evaluates the likelihood of different clustering settings. In our case, we use BIC to determine the number of clusters ($i.e.$, K). The $K$ of the maximum BIC is the optimal number of clusters. 
We pick the corresponding $K$ ($i.e.$, $K^{*}$) of the first decisive local maximum ($i.e.$, $BIC^{*}$).

\subsection{Measuring Reinforcement of User Interests}
\label{sec:user_rein}
We use cluster validity~\cite{Vazirgiannis2009} to compare the user embeddings of two user groups, and observe the changes in clustering through different months.
Originally, this technique is known as the procedure to evaluate how the clustering algorithm performs on the given datasets. The process evaluates the results on different parameter settings via a set of cluster validity indexes. 
These indexes for cluster validation can be grouped into two types, internal indexes (Section~\ref{sec:metric_internal}), and external indexes (Section~\ref{sec:metric_external}). The external indexes are based on the ground-truth clustering information, which is not always available for a dataset. On the contrary, the internal index can evaluate the clustering without knowing the optimal classification. We choose each of them to measure the temporal changes in clustering in both user groups.

\subsubsection{Internal Validity Indexes}
\mbox{}\\
\label{sec:metric_internal}
A good clustering algorithm is required to satisfy several valid properties, such as compactness, connectedness, and spatial separation~\cite{Julia2005computational}. 
One type of internal indexes is to evaluate to what extent the clusters satisfy these properties, and a prominent example is the Calinski-Harabasz index~\cite{calinski1974}.
Another type is applied to crisp clustering or fuzzy clustering~\cite{Pakhira2004ValidityIF,wang2007fuzzy}. 
Since we want to explore how user interests shift at the population level, 
we apply the former type of internal indexes on the clustering results of the user embeddings, in order to detect the polarization tendency in user preferences by tracking how the index changes over time.

\textbf{Calinski-Harabasz ($CH_{K}$)} index scores the clustering considering the variation ratio between the sum-of-squares between clusters ($SSB_{K}$) and the sum-of-squares within clusters ($SSW_{K}$) under $K$-class clustering. Based on this, we can compare the clustering of the same group at different times under the same setting, and a higher score indicates a better clustering. 
Let $N$ denote the size of the dataset $\{x_{1},\dots,x_{N}\}$, and 
$K$ denote the number of clusters. The centroids of clusters are denoted as $C_{i}$, $i=1,2,...,K$. For a data point $x_{j}$, it belongs to a cluster $p_{j}$ and we have the corresponding cluster centroid $C_{p_{j}}$, where $j=1,2,...,N$,$p_{j}=1,2,...,K$.
The Calinski-Harabasz index is thus calculated as follows:
\begin{equation}
\small
CH_{K} = { {{SSB_{K}}\over{SSW_{K}}} \cdot {{(N-K)}\over{(K-1)}}}
\end{equation}
where $SSW_{K} =\sum\limits_{j=1}^{N} \|x_{j}-c_{p_{j}}\|^{2}$, $SSB_{K} =\sum\limits_{i=1}^{K} \|c_{i}-\bar{X}\|^{2}$, and $\bar{X}$ represents the mean of the whole dataset.
Based on this definition, an ideal clustering result means that elements within a cluster congregate and elements in-between clusters disperse, leading to a high $CH_{K}$ value. 
Intuitively, we can assign users into clusters, and calculate the CH index for this clustering result based on the user embeddings. After a certain period of time, the user embeddings would change due to the user's new interactions during the time, and we can use the new embeddings to calculate the CH index without changing the users' cluster assignment. By comparing the CH index before and after the user embeddings change, we will be able to evaluate to what extent the user preferences have changed (see Figure~\ref{fig:ch}, details to be introduced later).
Furthermore, based on the user IDs, we can track how each user's preference changed in the latent space. 

\subsubsection{External Validity Indexes}
\mbox{} \\
\label{sec:metric_external}
We use external validity indexes to measure the similarity between the ``first block'' and the ``last block'' embeddings in terms of clustering.
This kind of indexes utilizes the ground truth class information to evaluate the clustering results. 
The clustering result close to the optimal clustering has high index scores.
In other words, the external indexes compute the similarity between the given clustering and the optimal clustering. 
 
External validity indexes are constructed on the basis of contingency tables ~\cite{wu2009external}. It is built as a matrix containing the interrelation between two partitions on a set of $N$ points. Partitions, $\textbf{P}=\{P_{1},P_{2},\dots,P_{K_{1}}\}$ of $K_{1}$ clusters and $\textbf{Q}=\{Q_{1},Q_{2},\dots,Q_{K_{2}}\}$ of $K_{2}$ clusters, give us a $K_{1}\times K_{2}$ contingency table ($C_{PQ}$) consisting the number of common points ($n_{ij}$) in $P_{i}$ and $Q_{j}$ as follows:
\begin{align}
\small
    C_{PQ}=\begin{bmatrix} n_{11} & n_{12} & \cdots & n_{1 K_{2}}  \\
 n_{21} & n_{22} &  \cdots  & n_{2K_{2}}  \\
 \vdots & \vdots & \ddots & \vdots  \\
   n_{K_{1}1} & n_{K_{1}2}  & \cdots & n_{K_{1}K_{2}}  \end{bmatrix}
\label{equ:contigency}
\end{align}
where $P_{i}$ and $Q_{j}$ are the clusters in $\textbf{P}$ and $\textbf{Q}$ with the size of $p_{i}$ and $q_{j}$, $i=1,2,\dots,K_{1}$, $j=1,2,\dots,K_{2}$. Therefore, we have  $p_{i}=\sum\limits_{j=1}^{K_{2}}n_{ij}$, $q_{j}=\sum\limits_{i=1}^{K_{1}}n_{ij}$,  and $\sum\limits_{i=1}^{K_{1}}\sum\limits_{j=1}^{K_{2}}n_{ij}=N$.

The techniques of comparing clusters for external validation are divided into three groups, pair-counting, set-matching, and information-theoretic~\cite{Vinh2009}. We use  \textbf{Adjusted Rand Index} (ARI)~\cite{Hubert1985}, which is a pair-counting index that counts the pairs of data points on which two clusters are identical or dissimilar. In this way, we can evaluate the portion of users shifting to another cluster on K-means clustering over time.
As a representative pair-counting based measure, ARI satisfies the three crucial properties of a clustering comparison measure~\cite{Vinh2009}: metric property, normalization, and constant baseline property.
It can be computed as follows:
\begin{equation}
\small
    \begin{aligned}
    ARI & =\frac{\sum\limits_{ij} {n_{ij}\choose 2}- [{\sum\limits_{i} {p_{i}\choose 2}}{\sum\limits_{j} {q_{j}\choose 2}}]/{N\choose 2}} {\frac{1}{2}[\sum\limits_{i} {p_{i}\choose 2} +\sum\limits_{j} {q_{j}\choose 2}]-[{\sum\limits_{i} {p_{i}\choose 2}}{\sum\limits_{j} {q_{j}\choose 2}}]/{N\choose 2}} 
    \end{aligned}
\end{equation}

ARI measures the similarity between two different clusterings (on two datasets) with the same clustering setting ($K$) for the same user group in our experiment. 
Traditionally, it evaluates the different clusterings under different clustering settings on the same dataset. 
The similarity between embeddings of the ``first block'' and the ``last block'' shows how much the clustering changes under the same $K$.
The higher the ARI score, the higher the similarity between the clusterings at the beginning and the end, and fewer changes in the latent space. 
The difference in ARI helps us to understand how the \textit{Following Group} acts under the influence of RS.
Since the distribution of user interest embeddings would change under the influence of external conditions, such as sales campaigns and the release of new products, the group that has fewer changes implies stable interests in certain types of items, showing the tendency of reinforcement.
Unlike the comparison in $CH_{K}$, a higher ARI implies that the preferences of the \textit{Following Group} is relatively reinforced.

\subsection{\mbox{Measuring the Changes of Content Diversity}}
\label{sec:metric_diversity}
To answer the second question, we measure the content diversity in recommendation lists. To detect how diverse a list of recommendations is, we compute the pairwise distance of item embeddings and use the average of the item distance to represent the content diversity of the list~\cite{nguyen2014exploring}.
We collect the first $N$ items recommended to users as the first recommendation list, and the last $N$ items as the last recommendation list. The Euclidean distance is computed between two item embeddings with the dimension of $D$ . Let $v_{i}$ be the vector of item embedding, $v_{i} \in R^{D}$, $i=1,\dots, N$, and $v_{i}^{d}$ is the value of $v_{i}$ on the $d$-$th$ dimension. Then we have the distance:
\begin{equation}
\small
    distance_{v_{i},v_{j}}=\sqrt{\sum\limits_{d=1}^{D}(v_{i}^{d}-v_{j}^{d})^{2}}
\end{equation}
The smaller the distance, the smaller the difference between the two items. We take the average of the pairwise Euclidean distance within the block to measure its content diversity~\cite{Ziegler2005}, and then utilize the temporal changes in content diversity to examine the effect of the recommender system on different user groups.

\section{Analyzing Echo Chamber}
\label{sec:result}
In this section, we first present the description of data after pre-processing in Section~\ref{sec:data_description}. 
Then, we present our clustering settings (i.e., the number of clusters selected) in Section~\ref{sec:result_settings}. We measure the cluster validity to examine the reinforcement in the \textit{Following Group} to answer RQ$1$ in Section~\ref{sec:result_validity}. 
Moreover, we evaluate the shifts in content diversity of recommendation lists so as to answer RQ$2$.
In the following, we refer to the user embedding based on clicked and purchased items as ``click embedding'' and ``purchase embedding'' respectively for simplicity.

\begin{table}[]
\vspace{-10pt}
\small
\begin{center}
\renewcommand{\multirowsetup}{\centering}
\begin{adjustbox}{max width=\linewidth}
\begin{tabular}{ ccc }
\toprule
Dataset & Statistic & Numerical Values \\ \midrule
& Num of click logs & $7,386,783$\\ 
Click Log& Num of users & $7,477$ \\
& Num of  items & $2,686,591$\\ \midrule
 & Num of purchase logs & $98,135$\\
Purchase Log& Num of users & $3,557$ \\
& Num of  items & $71,973$\\\midrule
\multirow{3}{*}{Browse Log}& Num of browse logs & $6,225,301$\\
 & Num of users & $7,417$\\
 & Num of items & $5,077,268$ \\
 \bottomrule
\end{tabular}
\end{adjustbox}
\end{center}
\caption{\label{font-table} Statistics of experiment data.}
\label{table:user_group}
\vspace{-30pt}
\end{table}

\subsection{Data Description}
\label{sec:data_description}
As what we have introduced in Section \ref{sec:data_prep}, experiments are performed on an extensive collection of real-world data with user click, purchase, and browse logs
. 
 The code of our entire experiment is released in our GitHub repository \footnote{https://github.com/szhaofelicia/EchoChamberInEcommerce}.
Meanwhile, after completing the data pre-processing in Section \ref{sec:data_prep}, $e.g.$, identifying recommendation followers and creating user-item interaction blocks, 
the detailed statistics for the final dataset regarding each type of user-item interaction are shown in Table~\ref{table:user_group}. 
Note that the \textit{Following Group} (i.e., recommendation takers) and the \textit{Ignoring Group} differ in size after the above pre-processing procedure. Since the size of the dataset would affect cluster analysis results, we need to resize the larger user group --- \textit{Following Group} --- in all types of user actions, to avoid the influence.
As a result, we sample from the \textit{Following Group} to make sure the \textit{Following Group} and the \textit{Ignoring Group} have an equal amount of users. 
We apply this operation (denoted as resize-sampling) to each kind of user interaction logs and generate three pairs of equal-sized user groups for our experiments. 

Additionally, we take a random sample of $p\%$ users ($p$-sampling) from the dataset in each computation, and repeat every experiment $50$ times to calculate the statistical significance for each measure.
We use $10\%$ for $p$-sample in Hopkins statistic, which is large enough to represent the distribution of a dataset.  
We use the ratio of $80\%$ for other indexes. We calculate the average of the 50 experiments as the final result. 
As a result, for the \textit{Following Group}, it takes two steps of samples -- resize-sampling and $p$-sampling -- before each computation. For instance, we need to compute cluster indexes for the \textit{Following Group} on the click logs $50$ times. For each time, we resize the \textit{Following Group} into $2452$ users and then take another sample of $80\%$. The final size would be $1962$ users, which is the same as the size of the \textit{Ignoring Group} after $p$-sample.


\begin{figure*}
\vspace{-15pt}
\centering
\subfigure[Click -- Following Group]{\includegraphics[width=0.267\textwidth]{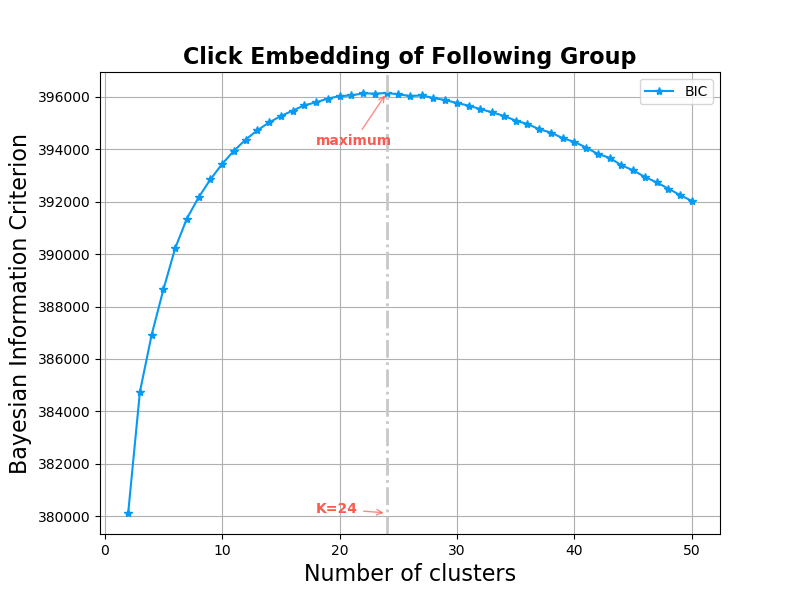}\label{fig:bic_a}}
\hspace{-16pt}
\subfigure[Click -- Ignoring Group]{\includegraphics[width=0.267\textwidth]{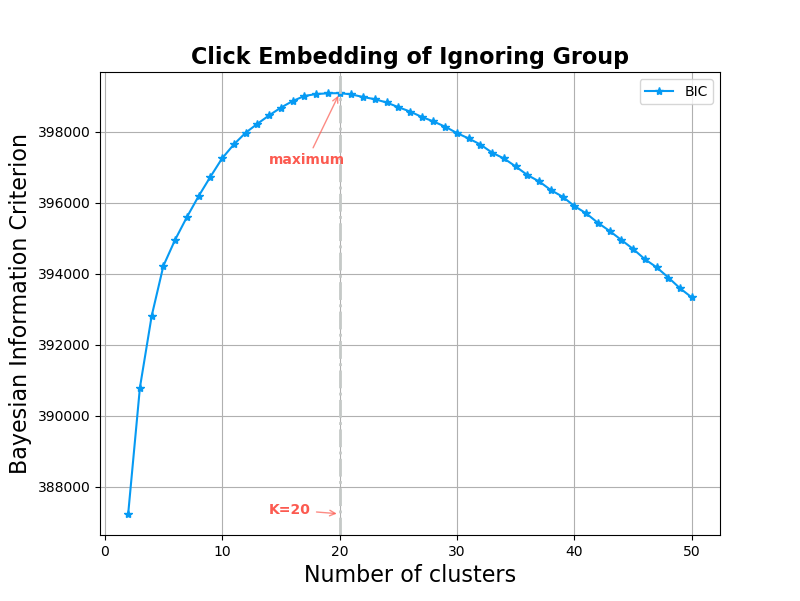}\label{fig:bic_b}}
\hspace{-16pt}
\subfigure[Purchase -- Following Group]{\includegraphics[width=0.267\textwidth]{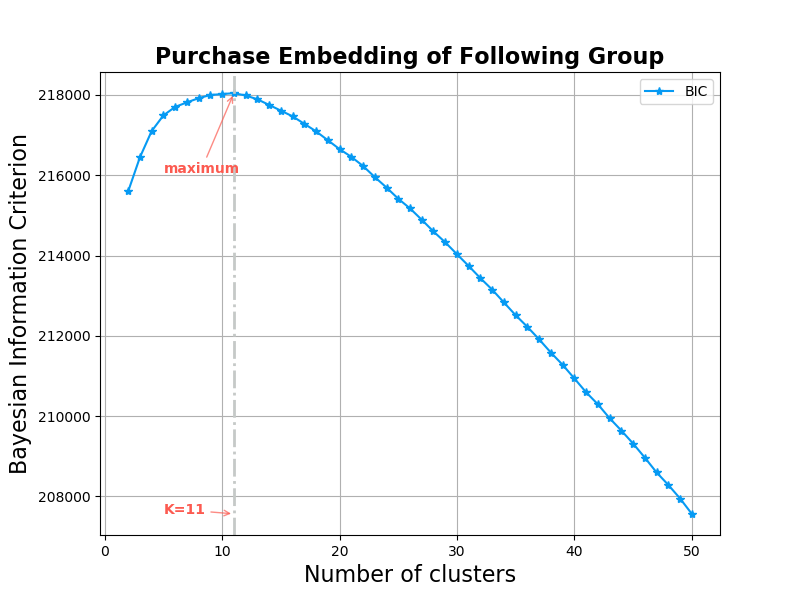}\label{fig:bic_c}}
\hspace{-16pt}
\subfigure[Purchase -- Ignoring Group]{\includegraphics[width=0.267\textwidth]{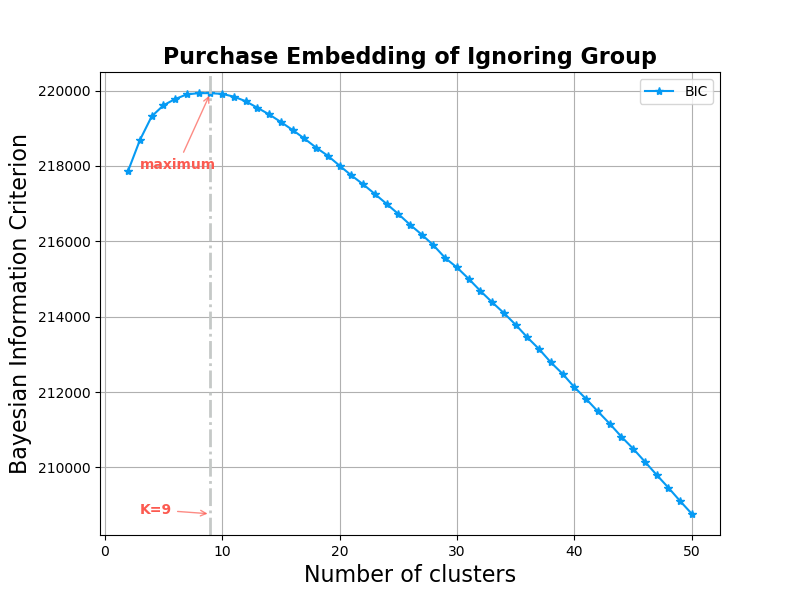}\label{fig:bic_d}}
\vspace{-10pt}
\caption{Bayesian Information Criterion (BIC). The $K$ of the maximum BIC is the optimal number of clusters.}
\label{fig:bic}
\end{figure*}

\subsection{Clustering Settings}
\label{sec:result_settings}
\subsubsection{Measuring the Cluster Tendency}
\label{sec:result_hopkins}

We examine the clustering tendency based on Hopkins statistic for the user embeddings of both \textit{Following Group} and \textit{Ignoring Group}, and the results are shown in Table~\ref{table:hopkins}. 
Our results show that both groups of user embeddings are clusterable ($H > 0.5$), so cluster analysis could generate meaningful results in the following experiments.
Meanwhile, we also observe that the clustering tendency for each group is not stable, and we utilize $p$-value of the $t$-test to examine its statistical significance since the temporal change of $H$ score is quite small.

Comparing the changes of $H$ score in the first and last block, the clustering tendency decreases in click embeddings 
but increases in purchase embeddings. 
Furthermore, the clustering tendency of purchase embedding is less changeable and even slightly increased because there are fewer local shifts in the latent space of purchase embedding compared with the temporal changes in click embedding.
This might attribute to the fact that 
users' tastes reflected in purchased items are relatively stable since users cannot choose to buy whatever they want as they need to pay the price for it. 

\begin{table}[t]
\centering
\resizebox{\linewidth}{!}{
\begin{tabular}{ cccccc }
\toprule
Action& User type & Amount & First Block & Last Block & P-value \\ \midrule
\multirow{4}{*}{Click} & All users & $4904$ & $0.7742$ & $0.7713$ & $4.33\mathrm{e}{-9}$\\ 
 & Following & $2452$ & $0.7756$ &$0.7746 $ &$3.05\mathrm{e}{-2}$ \\ 
& Ignoring & $2452$ & $0.7728$ &$0.7680 $ & $1.53\mathrm{e}{-21}$\\ 
 &Between-group p-value & $4904$ &$1.58\mathrm{e}{-8}$  & $2.88\mathrm{e}{-28}$ & \\ \midrule
\multirow{4}{*}{Purchase} & All users & $2916$ & $0.7264$ & $0.7279$ & $1.41\mathrm{e}{-2}$\\ 
 & Following & $1458$ & $0.7223$ &$0.7248 $ &$1.25\mathrm{e}{-6}$ \\ 
& Ignoring & $1458$ & $0.7305$ &$0.7310 $ & $0.27$\\ 
 &Between-group p-value & $2916$ &$2.65\mathrm{e}{-28}$  & $1.02\mathrm{e}{-24}$ & \\ \bottomrule
\end{tabular}}
\vspace{5pt}
\caption{Hopkins statistic. 
}
\vspace{-20pt}
\label{table:hopkins}
\end{table}

\begin{figure*}
\vspace{-15pt}
\centering
\subfigure[Click -- Following Group]{\includegraphics[width=0.267\textwidth]{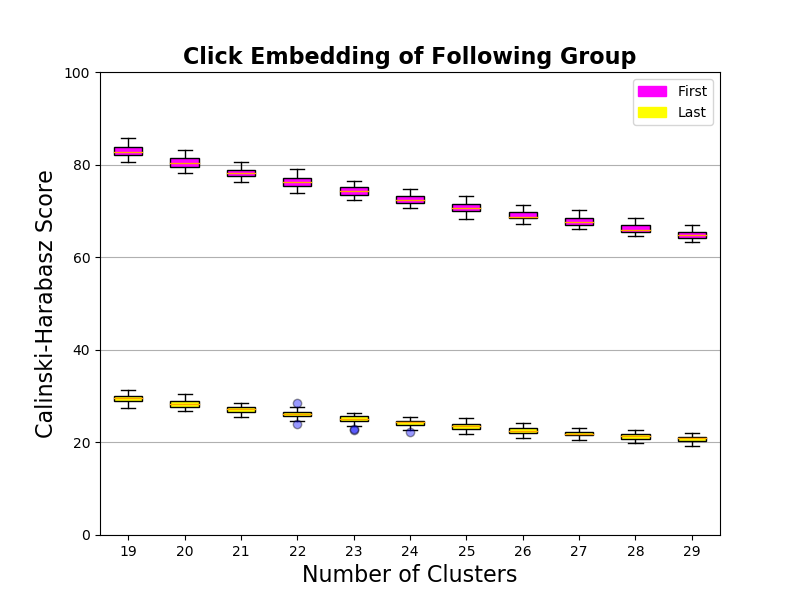}\vspace{-10pt}}
\hspace{-16pt}
\subfigure[Click -- Ignoring Group]{\includegraphics[width=0.267\textwidth]{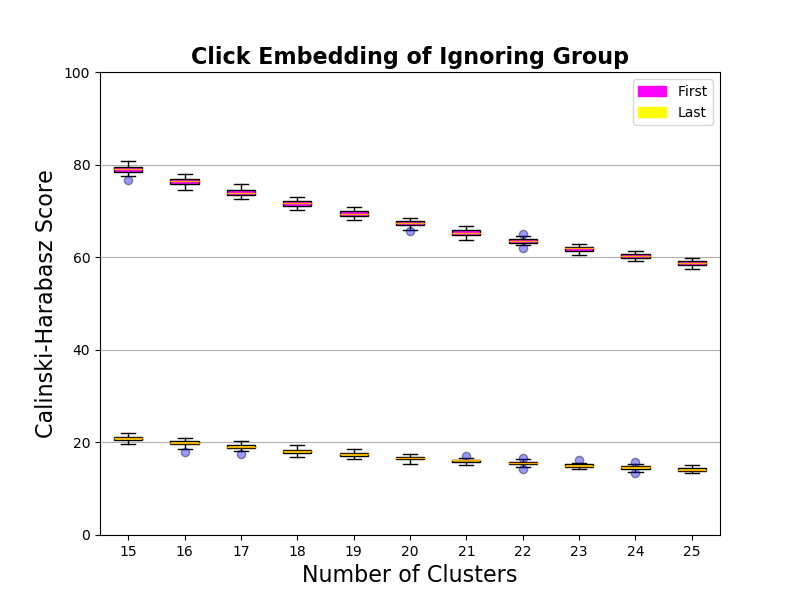}}
\hspace{-16pt}
\subfigure[Purchase -- Following Group]{\includegraphics[width=0.267\textwidth]{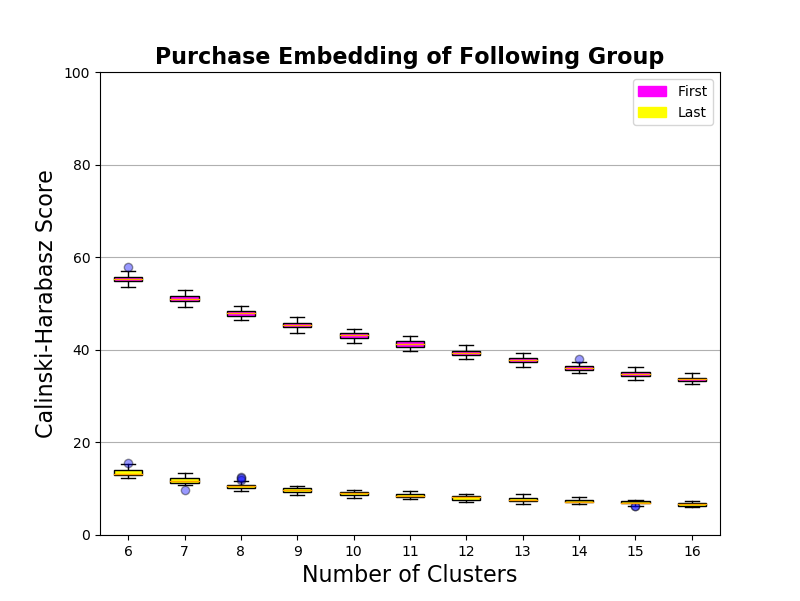}}
\hspace{-16pt}
\subfigure[Purchase -- Ignoring Group]{\includegraphics[width=0.267\textwidth]{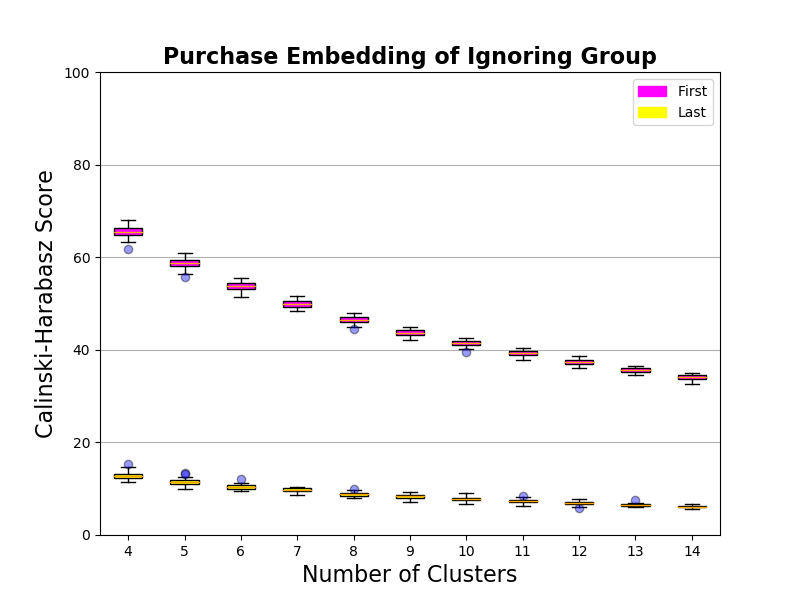}}
\vspace{-10pt}
\caption{CH \iffalse Calinski-Harabasz\fi score under different $K$ selected using BIC.}
\label{fig:ch}
\end{figure*}

\subsubsection{Select the Number of Clusters}
\label{sec:result_bic}
\mbox{}\\
Since we have shown that the \textit{Following Group} and \textit{Ignoring Groups} are clusterable in the previous section, we now use BIC to detect the optimal number of clusters ($K^{*}$) for each group of embeddings. 
The average BIC curves are plotted in Figure~\ref{fig:bic}. 
We do not force the clustering on each dataset to use the same number of clusters in consideration of underestimation caused by inappropriate $K$ settings. Clustering settings, such as $K$, have to fit the datasets well to guarantee the optimal clustering results. The inaccurate results might lead to overestimating or underestimating of the echo chamber effect.

We set the corresponding $K$ of the maximum BIC as the optimal number of clusters $K^{*}$. The $K^*$s is $24$, shown in Figure~\ref{fig:bic_a}, for the \textit{Following Group}, and $20$, shown in Figure~\ref{fig:bic_b}, for the \textit{Ignoring Group}, with the click embeddings. Meanwhile, $K^*$ is $11$, shown in Figure~\ref{fig:bic_c}, for the \textit{Followings Group}, and $9$, shown in Figure~\ref{fig:bic_d}, for the \textit{Ignoring Group}, with the purchase embeddings.  
Intuitively, we could directly compare the user groups with their own $K^{*}$, but
the local areas around maximum in the curves seem to be quite flat. As a result, we believe the measurements around $K^{*}$ would show more reliable and plausible results than the measurement exactly at $K^{*}$.
Thus, we execute the experiments in the range of $[K^{*}-5, K^{*}+5]$.
The average of results is used to examine the changes of clustering via cluster validity indexes introduced next.
Finally, the \textit{Following Group} uses $K$ with the range of $[19, 29]$ and $[6,16]$ for the click embedding and the purchase embedding respectively, and the \textit{Ignoring Group} uses $K$ with the range of $[15, 25]$ and $[4,14]$.

\vspace{-10pt}
\subsection{Results Analysis}
\label{sec:results}

\noindent 
\textbf{RQ$1$: Does RS reinforce user click or purchase interests?}
\label{sec:result_validity}
\vspace{-10pt}

\subsubsection{Internal Validity Indexes}
\label{sec:result_internal}
\mbox{} \\
As introduced in Section~\ref{sec:metric_internal}, CH can measure the extent of variation of the within-group clustering at different times. 
We first use CH to examine the tendency of reinforcement in user interests. The average scores of the CH index are plotted in Figure~\ref{fig:ch}, we can find that both groups have a drop in CH after three months at all $K$s, and CH also decreases as $K$ becomes larger both in the first blocks and the last blocks. The common decreasing trend over time in two groups might attribute to how we compute CH in the last blocks.

As we mentioned in Section~\ref{sec:metric_internal}, we assign the clustering partition results of the first blocks to the last blocks, which means that the same user will have the same cluster label both at the beginning and at the end.
However, the decrease in CH suggests that the temporal shifts in the user embeddings might have made the assignment of clusters unsuitable, i.e., the ideal clustering partition of the first blocks, can no longer serve as the ideal clustering partition of the last blocks due to the temporal shifts in the user embeddings.
Besides the effect of RS, the changes in user embeddings can also result from other factors. For instance, user interest can vary a lot, along with changes in external conditions in e-commerce, such as sales campaigns.
As a result, these temporal shifts of user embeddings in the latent space are reflected as the decrease in CH. 

In practice, we can hardly avoid this ``natural'' reduction caused by the e-commerce platform. As a result, we evaluate the difference between the two user groups to find the effect that comes from the RS. We compute the temporal decreases in CH for each group with $K$ in $[K^{*}-5, K^{*}+5]$ (see Table~\ref{table:ch}).
As is shown in the table, the drops of CH at $K^{*}$ are $48.22$ and $50.73$ for the \textit{Following Group} and \textit{Ignoring Group} in click embedding, and the average drops of CH in $[K^{*}-5, K^{*}+5]$ are $48.41$ and $50.95$ respectively. Moreover, purchase embeddings have similar results that the decreases of CH for the \textit{Following Group} are $32.74$ at $K^{*}$ and $33.26$ for average, and the reductions for the \textit{Ignoring Group} are $35.45$ and $37.29$. We further check the statistical significance of the difference between two groups, finding that all differences are at $95\%$ confidence interval (i.e., $p$-value is less than $0.05$). Overall, CH drops slower in \textit{Following Groups}, showing a more stable tendency than that in \textit{Ignoring Group}. Accordingly, the \textit{Ignoring Group}, which falls faster in CH,  disperses to a wide range on the latent space, reveals that it may receive a milder influence of reinforcement on user preference. 

The less dispersion of \textit{Following Group} in the latent space is possibly due to multiple factors. One the one hand, the \textit{Following Group} might have more users who hold on to the previous preference in items; on the other hand, \textit{Following Group} has fewer changes in their interests than the \textit{Ignoring Group} does. Either of the reasons could give us the conclusion that user interest in the \textit{Following Group} has a strengthening trend over time, resulting in that the dispersion in latent space is suppressed to some extent.

\begin{table}[t]
\centering
\renewcommand{\multirowsetup}{\centering}
\resizebox{\linewidth}{!}{
\begin{tabular}{ ccccccccc }
\toprule
& \multicolumn{3}{c}{Click} & \multicolumn{3}{c}{Purchase} \\ \cmidrule(lr){2-4} \cmidrule(lr){5-7}
 &Following &Ignoring  & P-value&Following &Ignoring & P-value\\ \hline
$k^{*}-5$ & $53.50$ & $58.14$ & $3.62\mathrm{e}{-41}$ & $41.76$ &$52.72$ & $6.94\mathrm{e}{-60}$ \\ 
$k^{*}-4$ & $52.21$ & $56.51$ & $2.10\mathrm{e}{-39}$& $39.30$ &$47.35$ & $3.54\mathrm{e}{-56}$ \\ 
$k^{*}-3$ & $51.12$ & $54.95$ & $1.58\mathrm{e}{-40}$ & $37.30$ &$43.41$ & $1.44\mathrm{e}{-50}$ \\ 
$k^{*}-2$ &$50.10$ & $53.54$ & $4.79\mathrm{e}{-35}$ & $35.60$ &$40.21$ & $3.50\mathrm{e}{-48}$ \\ 
$k^{*}-1$ & $49.19$ & $52.02$ & $7.64\mathrm{e}{-31}$& $34.13$ &$37.70$ & $7.47\mathrm{e}{-39}$ \\ 
$k^{*}$ & $48.22$ & $50.73$ & $4.29\mathrm{e}{-28}$ & $32.74$ &$35.45$ & $3.55\mathrm{e}{-29}$ \\ 
$k^{*}+1$ &$47.17$ & $49.38$ & $9.78\mathrm{e}{-22}$& $31.34$ &$33.62$ & $3.96\mathrm{e}{-24}$ \\ 
$k^{*}+2$ & $46.45$ & $48.00$ & $3.92\mathrm{e}{-15}$ & $30.09$ &$32.06$ & $8.70\mathrm{e}{-23}$ \\ 
$k^{*}+3$ &$45.74$ & $46.89$ & $8.67\mathrm{e}{-11}$ &$28.88$ &$30.54$ & $4.99\mathrm{e}{-19}$ \\ 
$k^{*}+4$ & $44.78$ & $45.69$ & $1.36\mathrm{e}{-7}$ &$27.78$ &$29.14$ & $2.69\mathrm{e}{-18}$ \\
$k^{*}+5$ &$44.05$ & $44.62$ & $2.61\mathrm{e}{-4}$ & $26.97$ &$28.00$ & $4.78\mathrm{e}{-14}$ \\ \midrule
AVE & $48.41$ & $50.95$ & $5.97\mathrm{e}{-32}$ & $33.26$ &$37.29$ & $1.06\mathrm{e}{-46}$ \\ \bottomrule
\end{tabular}}
\vspace{5pt}
\caption{Decreases in Calinski-Harabasz score. The corresponding $K$s for each groups are introduced in Section~\ref{sec:result_bic}}
\vspace{-15pt}
\label{table:ch}
\end{table}

\subsubsection{External Validity Indexes}
\label{sec:result_external}
\begin{figure*}
\vspace{-15pt}
\centering
\subfigure[Click -- Following Group]{\includegraphics[width=0.267\textwidth]{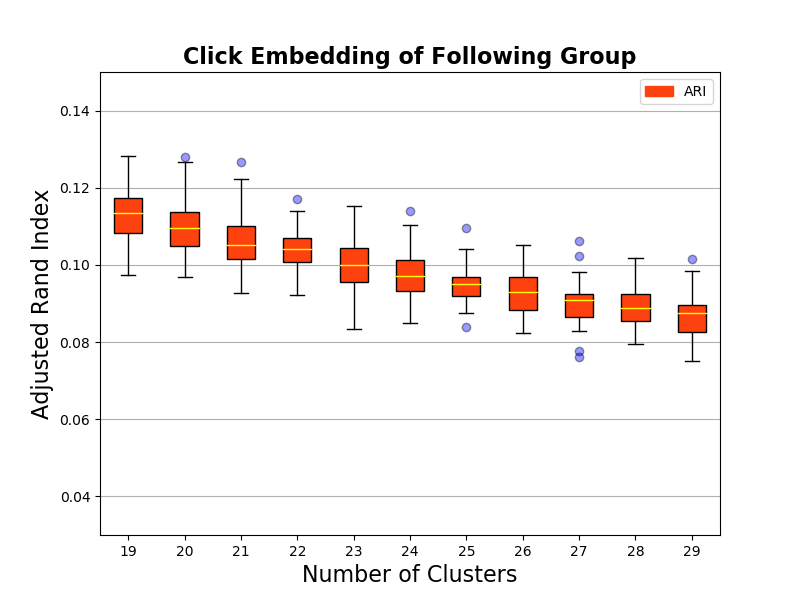}\label{fig:ari_a}}
\hspace{-16pt}
\subfigure[Click -- Ignoring Group]{\includegraphics[width=0.267\textwidth]{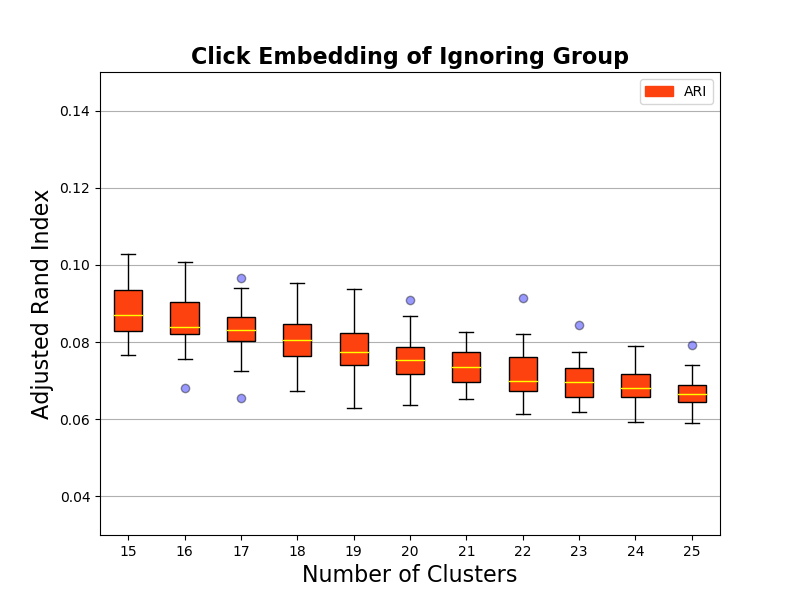}\label{fig:ari_b}}
\hspace{-16pt}
\subfigure[Purchase -- Following Group]{\includegraphics[width=0.267\textwidth]{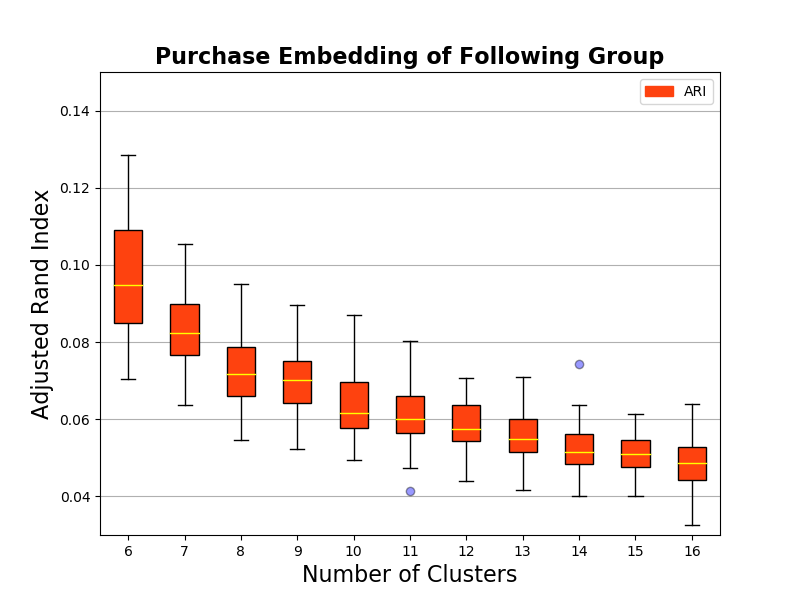}\label{fig:ari_c}}
\hspace{-16pt}
\subfigure[Purchase -- Ignoring Group]{\includegraphics[width=0.267\textwidth]{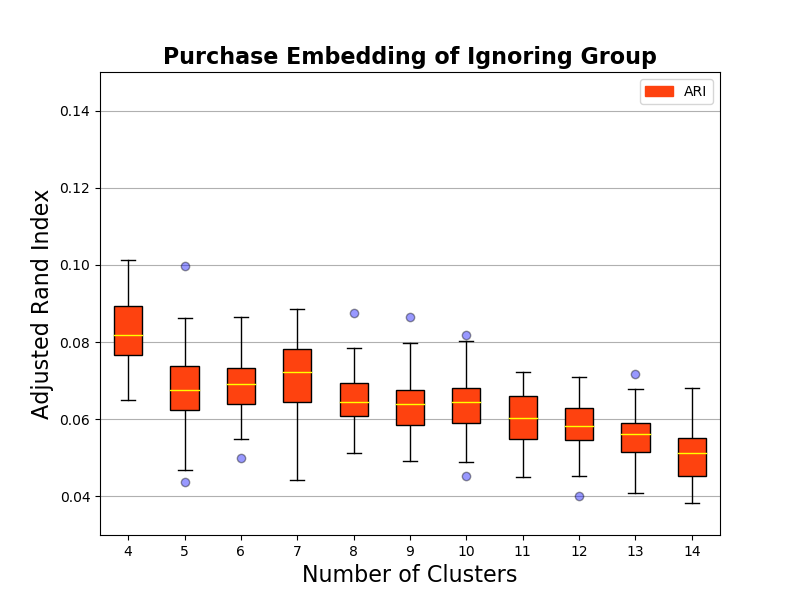}\label{fig:ari_d}}
\vspace{-10pt}
\caption{ARI \iffalse Adjusted Rand Index\fi under different $K$s, which are selected by BIC.}
\label{fig:ari}
\vspace{-5pt}
\end{figure*}

\mbox{} \\
Also, we examine the temporal changes in clustering via the external validity index, ARI. Unlike CH using the same labels on two datasets, ARI compares the different clusterings of the first and the last block and measures the similarity between clusterings. We plot the similarities (ARI) at different $K$s in Figure~\ref{fig:ari} and list the average results of ARI in Table~\ref{table:ari}. We find that in the click embedding, the \textit{Following Group} has a higher ARI than the \textit{Ignoring Group} (average ARI of $0.0986$ and $0.0765$ respectively with the $p$-value of $2.28\mathrm{e}{-51}$). In the purchase embedding, the difference between the two groups is not statistically significant; the $p$-value for the difference of the average ARI is $0.53$. Similar observation is also shown in the curves in Figure~\ref{fig:ari}, the curve in Figure~\ref{fig:ari_a} is higher than the curve in Figure~\ref{fig:ari_b}, but curves in Figure~\ref{fig:ari_c} and Figure~\ref{fig:ari_d} almost overlap. Let us take a look at each pair of ARI of different user groups in purchase embedding, the differences among half of the $K$s in $[K^{*}-5, K^{*}+5]$ are not significant. 

To sum up, the \textit{Following Group} has fewer temporal shifts in click embedding but no evident difference in purchase embedding. 
In other words, in terms of click interests, partitions at the beginning and the end in the \textit{Following Group} are more similar, indicating more connections. 
While in purchase interests, clustering at the end in both groups does not show the trend of sticking to the previous clustering.
More changes appearing in both groups in purchase embedding could be caused by the fact that users have fewer choices to purchase because of objective constraints, such as their incomes and the item prices. 
The effect of RS cannot ``force'' users to buy some items they cannot afford, thus, even if user interest has been reinforced, the shifts in preferences might not appear in purchase behaviors.
However, in click behaviors, the \textit{Following Group} seems to strengthen their preference under the effect of RS, since users are free to click items they are interested in, and their interests presented in click embedding do not have any other constraints. 
The group with higher ARI has fewer changes, which means they stick to the items they interacted with before and intensify their preferences.
The evidence confirms the conclusion in Section~\ref{sec:result_internal} that there exists the tendency of reinforcement in user interests in the \textit{Following Group}.

\begin{figure*}
\vspace{-10pt}
\centering
\subfigure[First Block -- Following Group]{\includegraphics[width=0.267\textwidth]{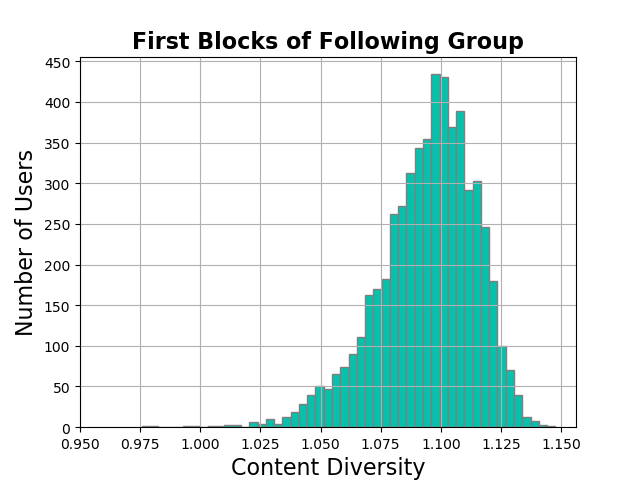}}
\hspace{-16pt}
\subfigure[Last Block -- Following Group]{\includegraphics[width=0.267\textwidth]{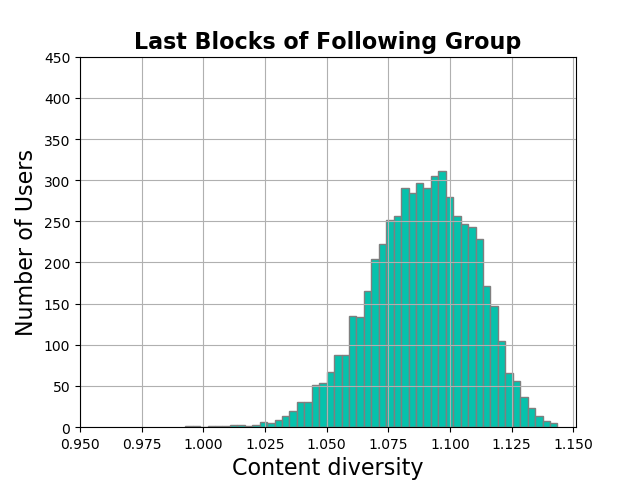}}
\hspace{-16pt}
\subfigure[First Block -- Ignoring Group]{\includegraphics[width=0.267\textwidth]{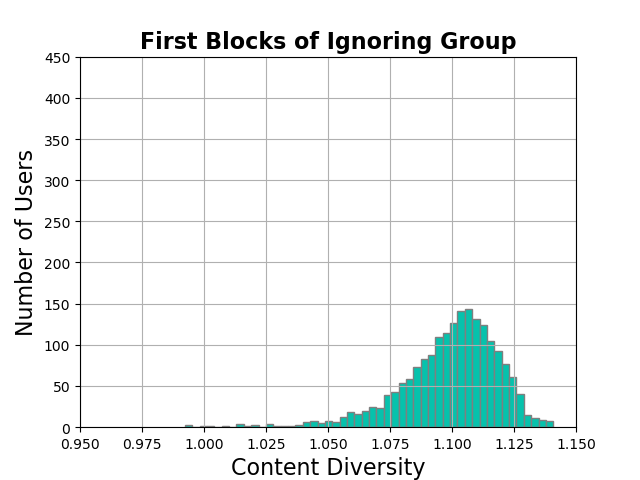}}
\hspace{-16pt}
\subfigure[Last Block -- Ignoring Group]{\includegraphics[width=0.267\textwidth]{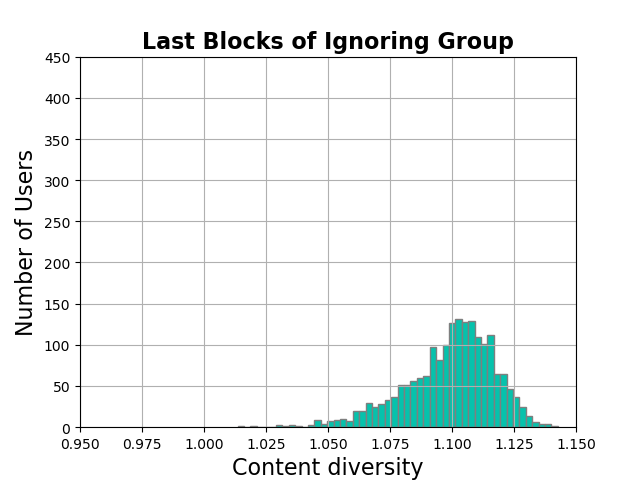}}
\vspace{-10pt}
\caption{Recommendation content diversity (pairwise distance between user embeddings) in two user groups.}
\label{fig:display}
\end{figure*}

\begin{table}[t]
\centering
\renewcommand{\multirowsetup}{\centering}
\resizebox{\linewidth}{!}{
\begin{tabular}{ccccccccc}
\toprule
& \multicolumn{3}{c}{Click} & \multicolumn{3}{c}{Purchase} \\ \cmidrule(lr){2-4} \cmidrule(lr){5-7}
 &Following &Ignoring  & P-value&Following &Ignoring & P-value\\ \hline
$k^{*}-5$ & $0.1136$ & $0.0877$ & $1.34\mathrm{e}{-33}$ & $0.0969$ &$0.0829$ & $3.70\mathrm{e}{-8}$ \\ 
$k^{*}-4$ & $0.1099$ & $0.0856$ & $1.85\mathrm{e}{-32}$& $0.0825$ &$0.0677$ & $1.67\mathrm{e}{-10}$ \\ 
$k^{*}-3$ & $0.1060$ & $0.0829$ & $1.23\mathrm{e}{-33}$ & $0.0725$ &$0.0686$ & $2.77\mathrm{e}{-2}$ \\ 
$k^{*}-2$ &$0.1040$ & $0.0811$ & $4.19\mathrm{e}{-34}$ & $0.0697$ &$0.0712$ & $0.35$ \\ 
$k^{*}-1$ & $0.0996$ & $0.0780$ & $2.88\mathrm{e}{-32}$& $0.0639$ &$0.0650$ & $0.52$ \\ 
$k^{*}$ & $0.0974$ & $0.0756$ & $5.52\mathrm{e}{-32}$ & $0.0615$ &$0.0635$ & $0.23$ \\ 
$k^{*}+1$ &$0.0949$ & $0.0734$ & $4.49\mathrm{e}{-41}$& $0.0585$ &$0.0634$ & $9.17\mathrm{e}{-4}$ \\ 
$k^{*}+2$ & $0.0929$ & $0.0717$ & $1.62\mathrm{e}{-33}$ & $0.0555$ &$0.0598$ & $4.12\mathrm{e}{-3}$ \\ 
$k^{*}+3$ &$0.0900$ & $0.0698$ & $1.13\mathrm{e}{-35}$ &$0.0524$ &$0.0583$ & $1.88\mathrm{e}{-5}$ \\ 
$k^{*}+4$ & $0.0890$ & $0.0687$ & $1.10\mathrm{e}{-39}$ &$0.0511$ &$0.0552$ & $1.48\mathrm{e}{-3}$ \\ 
$k^{*}+5$ &$0.0871$ & $0.0670$ & $4.20\mathrm{e}{-34}$ & $0.0489$ &$0.0510$ & $0.13$ \\ \midrule
AVE & $0.0986$ & $0.0765$ & $2.28\mathrm{e}{-51}$ & $0.0648$ &$0.0642$ & $0.53$ \\ \bottomrule
\end{tabular}}
\vspace{5pt}
\caption{ARI scores. Same as $CH_{K}$, the corresponding $K$s for each groups are introduced in Section~\ref{sec:result_bic}}
\vspace{-20pt}
\label{table:ari}
\end{table}

\vspace{5pt}
\noindent
\textbf{RQ$2$: If user interests are strengthened, is it caused by RS narrowing down the scope of items exposed to users?}
\label{sec:result_diversity}\\
After an affirmative answer to RQ$1$, we examine RQ$2$ to explore the potential cause of the reinforcement in user interests: narrowed contents offered to users.
To do so, we measure the content diversity of recommendation lists at the beginning and the end.
The distributions of content diversity (the average pairwise distance of item embeddings) in the first and last blocks for display are plotted in Figure~\ref{fig:display}, and the corresponding average content diversities are listed in Table~\ref{table:contentdiversity}
These distributions are approximately normal, and the first blocks have a larger density around higher content diversity than the last blocks do. 
Also, the distribution becomes dispersing over time, lowering the average of the whole group. Furthermore, the average content diversity gives us consistent observation. When paying attention to the overall temporal changes, we find that the content diversity among all users falls from $1.0960$ to $1.0937$ with $p$-value of $6.10\mathrm{e}{-11}$. Even though the drop is tiny, it indicates that both groups go through the trend of narrowing down the scope of the content displayed to users.   
Additionally, the content diversities in the \textit{Following Group} have a larger reduction from $1.0945$ to $1.0882$ 
than the reduction in \textit{Ignoring Group}. On the contrary, the decrease in the \textit{Ignoring Group} can even be ignored because of the high $p$-value of $0.67$. 
This is because RS learns more about followers from their interactions, such as click, purchase.
Thus, it is more likely for RS to provide items similar to what users have previously interacted with.

In e-commerce, user affects recommendations exposed to them through user actions, and their actions get influenced in return, these procedures form a feedback loop, which strengthens the personalized recommendation and shrinks the scope of the content offered to users. As a result, the filter bubble effect occurs in \textit{Following Group}. Conversely, \textit{Ignoring Group} only provide minimal information about their tastes in items, since they do not interact with RS much. Hence, RS recommends items from a broad scope to explore the users' preferences. The difference between the two groups is also statistically significant at all times. 
The \textit{Ignoring Group} has a higher diversity of $1.0992$ in the beginning, 
and   the difference is further enlarged in the end 
since the content diversity in \textit{Following Group} drops a lot. 
Then, the scope of items recommended to the \textit{Following Group} has been repeatedly narrowed down, strengthening user interests in this group as a consequence. 

As we claimed in answer to RQ$1$, the reinforcement in preferences --- echo chamber effect --- is reflected in the temporal shifts of user embeddings in clustering. Particularly, echo chamber appears in both user click interests and user purchase interests, but the effect in the latter is sort of slight. However, in other RS platforms, such as movie recommendations~\cite{nguyen2014exploring}, opposite observations appear in the \textit{Following Group}, indicating that RS helps users explore more items and mitigate the reduction in content diversity. 
One possible reason is that unlike products in e-commerce, promotional campaigns for movies mostly focus on those commercial films. Therefore, movie recommendation platforms could still fill the recommendation list with niche movies and slow down the reduction in content diversity.

\begin{table}[t]
\centering
\renewcommand{\multirowsetup}{\centering}
\resizebox{\linewidth}{!}{
\begin{tabular}{ ccccc }
\toprule
  & Amount & First & last & Within-group p-value \\ \midrule
All users &$3820$ &$1.0969$ & $1.0937$& $6.10\mathrm{e}{-11}$\\ \midrule
Following group  & $1910$ & $1.0945$ & $1.0882 $& $1.95\mathrm{e}{-20}$ \\ 
Ignoring group &  $1910$ & $1.0992$ &$1.0989 $ &$0.67$ \\ \midrule
 Between-group p-value & $3820$ &$2.13\mathrm{e}{-12}$  & $2.16\mathrm{e}{-56}$ & \\ \bottomrule
\end{tabular}}
\vspace{5pt}
\caption{\label{font-table} The content diversity of recommended items. 
}
\vspace{-20pt}
\label{table:contentdiversity}
\end{table}

\section{Conclusions and Future Work} \label{sec:conclusions}
In this paper, we examine and analyze echo chamber effect in a real-world e-commerce platform. We found that the tendency of echo chamber exists in personalized e-commerce RS in terms of user click behaviors, while on user purchase behaviors, this tendency is mitigated. 
We further analyzed the underlying reason for the observations and found that the feedback loop exists between users and RS, which means that the continuous narrowed exposure of items raised by personalized recommendation algorithms brings consistent content to the \textit{Following Group}, resulting in the echo chamber effect as a reinforcement of user interests. This is one of our first steps towards socially responsible AI in online e-commerce environments. Based on our observations and findings, in the future, we will develop refined e-commerce recommendation algorithms to mitigate the echo chamber effects, so as to benefit online users for more informed, effective, and friendly recommendations.

\bibliographystyle{ACM-Reference-Format}
\balance
\bibliography{reference}

\end{document}